\documentstyle[psfig]{article}
\begin{document}

\title{Binary coalescence of a strange star with a black
hole: Newtonian results.}

\author{William H. Lee$^{1}$, W\l odzimierz
Klu\'{z}niak$^{1,2,3}$ and Jon Nix$^{4}$}

\maketitle

$^{1}$Instituto de Astronom\'{\i}a, Universidad Nacional Aut\'{o}noma de
M\'{e}xico, Apdo. Postal 70--264, Cd. Universitaria, M\'{e}xico D.F. 04510
\\

$^{2}$Copernicus Astronomical Centre, ul. Bartycka 18, 00--716 Warszawa,
Poland \\

$^{3}$Institute of Astronomy, Zielona G\'ora University, ul. Lubuska 2,
65--265 Zielona G\'ora, Poland\\

$^{4}$Department of Physics, University of Chicago, 5640 S. Ellis
Ave., Chicago, IL 60637 \\

e-mail: wlee@astroscu.unam.mx, wlodek@camk.edu.pl,
jdnix@uchicago.edu \\

\abstract{We present Newtonian three--dimensional hydrodynamical
simulations of the merger of quark stars with black holes. The
initial conditions correspond to non-spinning stars in Keplerian
orbits, the code includes gravitational radiation reaction in the quadrupole
approximation for point masses. We find that the quark star is
disrupted, forming transient accretion structures around the black
hole, but 0.03 of the original stellar mass survives the
initial encounter and remains in an elongated orbit
as a rapidly rotating quark starlet, in all cases. 
 No resolvable amount of mass
is dynamically ejected during the encounters---the black hole
eventually
accretes $99.99\%\pm0.01\%$ of the quark matter initially present.}

\section{Introduction}

In this paper we study the binary coalescence of a black hole
and a quark star. Stellar population studies indicate that
if quark stars and black holes exist at all, such binaries
should exist in numbers significant from the point of view of
next-generation laser interferometric gravitational wave
detectors, but smaller than the number of Hulse-Taylor type binaries
(Belczy\'nski et al. 2001).
Coalescing quark stars also remain strong candidates for gamma-ray burst
sources (Paczy\'nski 1991, 2001; Haensel et al. 1991). 
This is reason enough to study such coalescences.
However, in this numerical study we address only one specific question:
how much quark matter, and with what velocities, is ejected when a quark 
star
coalesces with a black hole.
The interest here is in the speculations that such ejecta
may convert all neutron stars to quark stars.

\section{Quark stars, neutron stars, and the contamination problem}

Bodmer (1971) and Witten (1984)  
have conjectured that a three-flavor quark fluid is the ground state
of hadronic matter. This (up, down and) strange quark matter,
 if cold, would be stable in the bulk at zero pressure---spontaneous
 fission into individual hadrons or (hyper) nuclei would only occur
for the tiniest specks of quark matter,
 composed of less than a few thousand quarks each
(Farhi and Jaffe, 1984).
As Witten (1984) was quick to point out,
a sufficiently large (self-gravitating) quantity of quark matter
would be remarkably similar to conventional neutron stars---a
solar mass quark star would be about 10 km across, and the maximum
mass of a quark star stable against collapse to a black hole
would be about $2M_\odot$. The TOV equation for quark stars of masses up 
to
the maximum  value has also been solved by Itoh (1970);
Brecher and Caporaso (1976); Haensel, Zdunik and Schaeffer (1986);
as well as by Alcock, Farhi and Olinto (1986a), who also give a detailed
discussion of the properties of these ``strange'' stars and of the
astrophysical context. For recent reviews see
Cheng, Dai and Lu (1998) and Madsen (1999).

The existence of self-bound quark matter and of quark stars
remains a hypothesis. Even so, we are now witnessing a revival
in their theoretical study, prompted no doubt by advances
in computational techniques and in the maturing of
X-ray and gamma-ray astronomy,
as well as by a deeper understanding of collective effects
in quark matter (e.g., Alford, Rajagopal and Wilczek 1998;
Rapp et al. 1998).

Several groups have computed the
structure of rotating quark stars  in full general relativity,
and discussed their external
metric in the context of kHz quasi-periodic oscillations (QPOs)
observed in certain X-ray binaries (Gourgoulhon et al. 1999,
Stergioulas et al. 1999, Gondek-Rosi\'nska et al. 2001,
Bombaci et al. 2000), and an even greater number
have investigated the possible
connection between quark stars and energetic phenomena
such as gamma-ray bursts (Paczy\'nski 1991;
Haensel, Paczy\'nski and Amsterdamski 1991; Cheng and Dai 1996),
soft gamma-repeaters, a.k.a. SGRs (Alcock, Farhi and Olinto 1986b;
Horvath et al. 1993;
Cheng and Dai 1998, 2002; Dai and Lu 1998; 
Zhang, Xu and Qiao 2000; Usov 2001),
and radio pulsars (Xu et al. 1999).
 This reference list is far from exhaustive.

However, there is a shadow over all this activity. The Galaxy must not be
contaminated with ``seeds'' of quark matter. Such seeds present no
danger to the Earth, because ions are repelled
 by the high ($\sim 10\,$MeV) Coulomb barrier surrounding quark matter
of the type discussed here (Farhi and Jaffe 1984; Alcock et al. 1986a).
 But neutrons are easily absorbed and
dissolved by quark matter,
and Witten (1984) noted that if a coalescing binary
of the Hulse-Taylor type contains a quark star, and if Clark and Eardley's
(1977) speculation on the amount of matter ejected 
(said to be $\sim 0.1M_\odot$) is correct, then ``there could be
$10^5M_\odot$ of quark matter free in our Galaxy'' and even the
smallest bit of quark matter inside a neutron star
would convert it to a quark star by absorbing all neutrons.
On the other hand, it has been argued that some radio pulsars
are neutron stars and cannot be quark stars
 (Alpar 1987, Madsen 2000).
This would place an extremely severe limit on the
space density of quark nuggets, a limit thought to be
drastically violated by the expected ejection of matter in 
those coalescing binaries
in which at least one of the components is a quark star,
and the other is equally or more compact
(Madsen 1988, Caldwell and Friedman 1991).

\section{Aim of this study}

We would like to find out if binary coalescence events
do indeed pose a problem for the hypothethical co-existence of quark
stars and neutron stars in our Galaxy.
There are several aspects to this issue.

The first and non-controversial statement is that if quark matter is 
stable,
then even the smallest quark seed present in the interior
of a neutron star would convert it to a quark star 
(Witten 1984, Olinto 1987).
Such a seed may arise  spontaneously when neutron-star matter exceeds
a certain density, as in the center of neutron stars or in supernovae,
or it may be captured from the ambient medium by a neutron star
or its massive stellar progenitor (Alcock et al. 1986a). It is the
latter possibility that concerns us here.

The second question is whether neutron stars and quark stars 
actually co-exist in our Galaxy.
If quark fluid is not the most stable form of hadronic matter,
the issue is moot. Ditto, if all the presumed neutron stars
 (radio pulsars, X-ray bursters, etc.) are in fact quark stars.
The question really arises only if at least some of the observed
compact objects are neutron stars. It has been argued that
glitching radio pulsars (such as the Crab and Vela) are really
conventional neutron stars (Alpar 1987)---the observed
occasional impulsive changes
in the period of such young radio pulsars
have been understood in terms of redistribution of angular momentum
at the base of the crust of neutron stars, and quark stars
are unable  to support a crust of the requisite moment of inertia
(Alcock et al. 1986a). It has also been argued that the r-mode
instability limits the rotation rate of quark stars, at least for some
forms of quark matter, to values far below the ones observed
in millisecond pulsars (Madsen 2000).
These arguments would lead to very stringent limits on the
space density of quark matter, although it has also been argued
that recent work on the crystalline
phase of superfluid quark matter may alleviate these concerns
(Alford et al. 2000).

The third and fourth questions relate to the total mass
of quark matter dispersed in the Galaxy and to the size distribution
of its fragments.
It is generally {\sl assumed} that about $0.1M_\odot$ of quark
matter is ejected in a coalescing quark star binary.
Madsen (1988) finds  that if the ejected matter
is dispersed into small nuggets of baryon number $A<10^{28}$,
i.e., less than (possibly much less than) 10 kg in mass,
then such a nugget can be captured by a pre-supernova
star and come to rest at the center of its core.
As discussed above, this would  lead to the conversion of any neutron
 star,
subsequently formed in the supernova, to a quark star.
Madsen then concludes that even one coalescence event ejecting
 quark matter
would be enough to seed all pre-supernova stars.
Caldwell and Friedman (1991) specifically discuss
the fate of quark matter fragments formed in the disruption of a quark
star coalescing with a black hole, and come to the same conclusions.

The disrupted star has a mass exceeding $10^{30}\,$kg,
i.e., $A\sim10^{57}$.
No numerical simulation is currently capable of having a dynamic
range of 29 orders of magnitude, so it is unrealistic to expect
the  hydrodynamic computation reported here to settle the issue of the size
of the droplets into which quark matter breaks up in the coalescing
binary. But it is possible, and this is the main aim of the present study,
to set upper limits to the total amount of quark matter ejected
in the form of droplets. 
We have previously successfully performed simulations of coalescing
``neutron star'' binaries
allowing a mass resolution down to $10^{-5}M_\odot$
to be achieved on desk-top work-stations
(Lee and Klu\'zniak 1995; Klu\'zniak and Lee 1998; and papers I,II,III,IV:
Lee and Klu\'zniak 1999a,b, Lee 2000, 2001, respectively).
This may not seem to be very constraining, as Caldwell and
Friedman (1991) argue that after $10^6$ coalescences of strange--black
hole systems ``the average strangelet content exceeds by forty-one orders
of magnitude the minimum density needed to seed the star.''
But in fact, the difference between between $10^{-1} M_\odot$
and  $10^{-3} M_\odot$ may be critical, because forty-four of the
forty-one orders of magnitude come from the assumed fragmentation
of quark matter, supposed to be occurring close to the black hole.
In our Newtonian simulations only a small amount of matter 
($\sim10^{-2}M_\odot$)
never approaches the black hole, and if it is a fraction of this
matter that is ejected, the fragmentation into small quark nuggets
may never have occurred.

If it is true, as we find, that Clark and Eardley's guess of
$0.1M_\odot$ for the mass ejected in the coalescence of two
{\it neutron} stars---a value uncritically adopted
(also by Klu\'zniak, 1994) in all estimates of the ambient density
of quark nuggets in the Galaxy---is not supported
by actual hydrodynamic simulations of the coalescence of 
a {\it quark} star
and a black hole,  astrophysical arguments against the existence of
stable quark matter at zero pressure will have to be re-examined.

Much of the argument for dispersal of quark nuggets in the Galaxy
rests on the assumption of violent collisions between quark
fragments, prior to their ejection from the binary.
A subsidiary aim of this paper is to check whether the coalescence
process is indeed conducive to fragmentation of quark matter
into fragments of low baryon number.
 
\section{The choice of simulation}

In this paper we report the results of a Newtonian SPH study 
of coalescing 
binaries in which one component is a massive quark star and the
other a black hole, about twice or three times as massive.
To our knowledge, this is the first 3-d hydrodynamical study to 
be performed
of coalescing quark stars. We have chosen to first study a black hole
as the second component in the binary rather than
a quark star,  for four reasons, giving us some measure of confidence
that the results reported may be qualitatively correct.
First, the coalescence process is essentially over much more quickly
when one of the components is a black hole, allowing more simulations
to be carried out in a given time. This is both because
the actual physical process is shorter and because the number of SPH
particles  decreases during the simulation as they are swallowed
by the black hole.
Second,
all the SPH particles in this simulation model a single quark
star, instead of two. This gives a substantial gain in the resolution.
Third, in the Newtonian approach used here, it is easier to simulate
some qualitative features of general relativity in black-hole
accretion than in accretion
onto material stars. 
The only qualitative effect of general relativity
modeled in the present  study is the irreversible accretion of matter
by the black hole---the Paczy\'nski-Wiita potential has been
widely used in other studies of  accretion disk
to mock up the presence of the
marginally stable orbit, but the simulations reported here
 relied on a fully Newtonian potential.
Fourth, unlike in the case of two coalescing non-rotating stars,
no vortex sheet is expected to form in a black hole binary.

\section{Numerical Method}

For the calculations presented in this paper, we have used the method
known as Smooth Particle Hydrodynamics (SPH) (see Monaghan~1992 for a
review of the method). The code is essentially the same one used
previously to model the coalescence of black holes with polytropic
stars (papers I--IV). 
The equations of motion include an artificial viscosity
term, to handle the presence of shocks and avoid particle
interpenetration. The standard form (see Monaghan~1992) includes terms
both for shear and bulk viscosity. During dynamical simulations of
coalescing binaries, accretion disks are often formed, and thus the
effects of a shear viscosity can have a substantial impact on their
evolution. To minimize this effect, we have used the artificial
viscosity prescription proposed by Balsara (1995). The momentum and
energy equations for a given SPH particle are then given by:
\begin{equation}
\frac{d\vec{v}_{i}}{dt}=-\sum_{j}m_{j}\left(
\frac{2\sqrt{P_{i}P_{j}}}{\rho_{i}\rho_{j}}+
\Pi_{ij}\right)\nabla_{i}W_{ij}-\nabla\Phi_{i}+\vec{a}_{i}^{RR},
\end{equation}
and 
\begin{equation}
\frac{du_{i}}{dt}=\frac{1}{2}\sum_{j}m_{j}\left(
\frac{2\sqrt{P_{i}P_{j}}}{\rho_{i}\rho_{j}}+
\Pi_{ij}\right)(\vec{v}_{i}-\vec{v}_{j})\cdot\nabla_{i}W_{ij}.
\end{equation}
Here $\Pi_{ij}$ is the artificial viscosity term and $\vec{v}$, $P$,
$u$, $\Phi$, $a^{RR}$ and $W$ are the velocity, pressure, internal
energy per unit mass, gravitational potential (which includes the self
gravity of the fluid as well as the contribution arising from the
presence of the black hole), gravitational radiation reaction
acceleration, and the smoothing kernel respectively. For the kernel we
use the spline form of Monaghan \& Lattanzio (1985). The viscous term
is given by
\begin{equation}
\Pi_{ij}=\left( \frac{P_{i}}{\rho^{2}_{i}} +
\frac{P_{j}}{\rho^{2}_{j}}\right)(-\alpha \mu +\beta \mu^{2}_{ij})
\end{equation} 
where
\begin{eqnarray*}
\mu_{ij}=\left \{ \begin{array}{ll} \frac{(\mbox{\boldmath
$v$}_{i}-\mbox{\boldmath $v$}_{j}) \cdot (\mbox{\boldmath
$r$}_{i}-\mbox{\boldmath $r$}_{j})}{h_{ij}(|\mbox{\boldmath
$r$}_{i}-\mbox{\boldmath $r$}_{j}|^{2}/h_{ij}^{2})+\eta^{2}}
\frac{f_{i}+f_{j}}{2c_{ij}}, & (\mbox{\boldmath
$v$}_{i}-\mbox{\boldmath $v$}_{j}) \cdot (\mbox{\boldmath
$r$}_{i}-\mbox{\boldmath $r$}_{j}) <0 \\ 0, & (\mbox{\boldmath
$v$}_{i}-\mbox{\boldmath $v$}_{j}) \cdot (\mbox{\boldmath
$r$}_{i}-\mbox{\boldmath $r$}_{j}) \geq 0
\end{array} \right.
\end{eqnarray*}
and $f_{i}$ is the form-function for particle {\em i} defined by
\begin{eqnarray*}
f_{i}=\frac{|\mbox{\boldmath $\nabla$} \cdot \mbox{\boldmath
$v$}|_{i}}{|\mbox{\boldmath $\nabla$} \cdot \mbox{\boldmath
$v$}|_{i}+|\mbox{\boldmath $\nabla$} \times \mbox{\boldmath
$v$}|_{i}+\eta'c_{i}/h_{i}}.
\end{eqnarray*}
The factor $\eta'\simeq 10^{-4}$ in the denominator prevents numerical
divergences. The sound speed at the location of particle {\em i} is
denoted by $c_{i}$, and $\alpha$ and $\beta$ are constants of order
unity. The divergence and curl of the velocity field are evaluated
through
\begin{eqnarray*}
(\mbox{\boldmath $\nabla$} \cdot \mbox{\boldmath
$v$})_{i}=\frac{1}{\rho_{i}}\sum_{j}m_{j}(\mbox{\boldmath
$v$}_{j}-\mbox{\boldmath $v$}_{i})\cdot \mbox{\boldmath
$\nabla$}_{i}W_{ij}
\end{eqnarray*}
and
\begin{eqnarray*}
(\mbox{\boldmath $\nabla$} \times \mbox{\boldmath
$v$})_{i}=\frac{1}{\rho_{i}}\sum_{j}m_{j}(\mbox{\boldmath
$v$}_{j}-\mbox{\boldmath $v$}_{i})\times \mbox{\boldmath
$\nabla$}_{i}W_{ij}.
\end{eqnarray*}
This form of the viscosity vanishes in regions of strong vorticity,
when $\nabla \times \vec{v} \gg \nabla \cdot \vec{v}$, but remains in
effect if compression dominates in the flow ($\nabla \cdot \vec{v} \gg
\nabla \times \vec{v}$). This allows us to minimize the effects of
artificial viscosity on the evolution of disk--like structures
in the simulations, when they appear.

In this study, unlike before, we take the gravitational acceleration
of a volume of fluid to be proportional to its total energy density,
i.e., we reinterpret $\rho$ in all the above
equations (but not in the self-gravity term implicit in $\Phi_i$)
as the energy density
divided by $c^2$, and we add any changes in the internal energy
(eq. [2]) to $\rho c^2$.
To model quark matter we use the simplest
MIT equation of state (e.o.s.), where the pressure is given by
$P=c^{2}(\rho-\rho_{0})/3$ for $\rho >\rho_{0}$, and is zero
otherwise. Note that for $\rho\le\rho_0$ the viscous stress vanishes (eq. [3]),
and when the radiation reaction is turned off as well the equation
of motion (1) is that of dust.

In compact binaries, the orbital decay is driven primarily by the
emission of gravitational waves. To take this effect into account, we
include the back reaction on the system, computed
in the quadrupole approximation for point masses (see e.g. Landau \&
Lifshitz~1975), so that the rates of energy and angular momentum loss
are given respectively by
\begin{equation}
\frac{dE}{dt}=-\frac{32}{5} \frac{G^{4}(M_{\rm SS}+M_{\rm BH})
(M_{\rm SS}M_{\rm BH})^{2}}{(cr)^{5}} 
\label{eq:dedt}
\end{equation}
and
\begin{equation}
\frac{dJ}{dt}=-\frac{32}{5 c^{5}} \frac{G^{7/2}}{r^{7/2}}
M_{\rm BH}^{2} M_{\rm SS}^{2} \sqrt{M_{\rm BH}+M_{\rm SS}},
\label{eq:djdt}
\end{equation}
where $r$ is the binary separation (defined as the distance between
the black hole and the center of mass of the strange star).

The corresponding acceleration on each binary component is then given
by:
\begin{eqnarray}
\mbox{\boldmath $a$}^{*}=-\frac{1}{q(M_{\rm SS}+M_{\rm BH})} 
\frac{dE}{dt}
\frac{\mbox{\boldmath $v$}^{*}}{(v^{*})^{2}} 
\label{eq:reaction}\\
\mbox{\boldmath $a$}^{\rm BH}=-\frac{q}{M_{\rm SS}+M_{\rm BH}} \frac{dE}{dt}
\frac{\mbox{\boldmath $v$}^{\rm BH}}{(v^{\rm BH})^{2}}
\label{eq:reactionbh}
\end{eqnarray}
where $v^{*}$ is the velocity of the quark star and $v^{\rm BH}$
that of the black hole, $q$ is the mass ratio of the components.

The application of the above equations is trivial in the case of the
black hole, since we always treat it as a point mass. For the star, we
apply the same acceleration to each SPH fluid particle, using
equation~(\ref{eq:reaction}) evaluated at the center of mass of the
fluid, so that we have:
\begin{eqnarray}
\mbox{\boldmath $a$}^{i}=-\frac{1}{q(M_{\rm SS}+M_{\rm BH})} \frac{dE}{dt}
\frac{\mbox{\boldmath $v$}^{*}_{cm}}{(v^{*}_{cm})^{2}}.
\label{eq:reactionsph}
\end{eqnarray}
Once the star is tidally disrupted, this approximation clearly becomes
meaningless, and so we switch off the corresponding terms when the
binary separation becomes smaller than the tidal disruption radius
$r_{tidal}=CR_{\rm SS}(M_{\rm BH}/M_{\rm SS})^{1/3}$,
 where $C$ is a constant
of order unity. This formulation of gravitational radiation
back--reaction has been used before for coalescing compact binaries
(e.g. Davies et al.~1994; Rosswog et al.~1999; Lee \& Klu\'zniak~1999b).

\section{Initial conditions}

We initially construct a spherical star by placing $N$ particles of
equal mass on a cubic three--dimensional grid and including a damping
term in the equations of motion for an isolated star. The system then
relaxes for approximately twenty freefall times ($t_{ff}\approx
(G\overline{\rho})^{-1/2}$). Table~\ref{IC} shows the initial
parameters used for our dynamical runs. We have used three different
values for the initial mass of the strange star, corresponding to the
maximum mass for a given value of $\rho_{0}$ 
(as noted by Witten~[1984], $M_{\rm max} \propto \rho_{0}^{-1/2}$;
for a discussion of physical bounds
on $M_{\rm max}$ see Zdunik~et~al.~[2000]).
The black hole is modeled as a spherical
vacuum cleaner---a point mass
producing a Newtonian potential $\Phi=-GM_{BH}/r$, with an absorbing
boundary at the Schwarzschild radius $r_{Sch}=2GM_{BH}/c^{2}$.
The
mass ratio is defined as $q=M_{\rm SS}/M_{\rm BH}$. For each value of
$M_{SS}$ we have performed calculations for two different values of
$q$, giving a total of six dynamical runs, shown in Table~\ref{IC}.

\begin{table}
\caption{Initial conditions ($N=17 256$ for all runs)}
\label{IC}
\begin{center}
\begin{tabular}{ccccccccc}
& & & & & & \\
Run & ${\rho_{0}}$
& $M_{\rm SS}$ & $R_{\rm SS}$ & $t_{ff}$ & $q$ 
& $r_{i}$ & $\nu_{orb}$ \\
~&$[10^{14}$g$\,$cm$^{-3}]$ &$[M_{\odot}]$  &[km]& [ms] & &
$[R_{\rm SS}]$ & [Hz]&\\
A & 7.318 & 1.5 &  9.0 & 0.06 & 0.5 & 3.25 & 775.59 \\
B & 7.318 & 1.5 &  9.0 & 0.06 & 0.3 & 3.70 & 767.37 \\
C & 4.116 & 2.0 & 12.0 & 0.08 & 0.5 & 3.25 & 581.69 \\
D & 4.116 & 2.0 & 12.0 & 0.08 & 0.3 & 3.70 & 575.53 \\
E & 2.634 & 2.5 & 15.0 & 0.10 & 0.5 & 3.25 & 465.35 \\
F & 2.634 & 2.5 & 15.0 & 0.10 & 0.3 & 3.70 & 460.42 \\
\end{tabular}
\end{center}
\end{table}

To perform the dynamical simulations (described below), we place the
star a distance $r_{i}$ from the black hole and give the binary
components the azimuthal velocity corresponding to a Keplerian binary
with angular velocity $2\pi\nu_{\rm orb}=\sqrt{G(M_{\rm
SS}+M_{BH})/r^{3}}$, plus the radial velocity corresponding to
point--mass inspiral.

 Every SPH particle in
the star is given the same azimuthal velocity, and thus the system
corresponds to one in which the star is not spinning in an external
(inertial) frame of reference.
We have two reasons for choosing a non-spinning star
at the beginning of the run. This initial condition is believed to
be realistic since the shear viscosity of quark matter is believed to
be smaller than in neutron-star matter, and in neutron stars
tidal synchronization can be neglected
(Kochanek~1992; Bildsten \& Cutler~1992).
Further, past experience
(papers I through IV)
teaches us that the ejection of matter from 
tidally locked polytropes is much smaller than from non-spinning
polytropes (for which the coalescence process is much more violent).
The present simulations of quark-star coalescence resemble to a certain extent
our earlier simulations for stiff polytropes, hence we expect the
same dependence to hold here.
 A non-spinning quark star is the right choice for the
initial conditions, if we are  to place secure upper bounds on
the amount of matter ejected from the binary.

As it turns out,
at the start of the dynamical calculation, a tidal bulge appears on
the star, in the direction facing the black hole. This is simply because
the initial configuration (i.e., spherical star plus
point-mass companion) is not in equilibrium at $t=0$. Thereafter the
star spirals in due to gravitational radiation reaction (at the
initial binary separations given in Table~\ref{IC}, the decay
timescale due to the emission of gravitational waves is comparable to
the orbital period).
In trial runs we have placed the star also at various larger
initial separations. The outcome of the coalescence was found to be
insensitive to the choice of $r_i$.

\section{Comparison with previous simulations}
We are in a position to compare the results presented here for
quark-matter e.o.s., $P=c^{2}(\rho-\rho_{0})/3$, with those obtained
for the polytropic e.o.s. $P=K\rho_b^\Gamma$
or $P=(\Gamma-1)\rho_bu$ (here $\rho_b$ denotes the baryon
rest-mass density). We had previously carried out coalescence
simulations, with the same code as the one used here,
for stiff polytropes with $\Gamma=3$ (papers I, III), and
for soft polytropes with  $\Gamma=5/3$ and  $\Gamma=2$ (papers II, IV).
In all cases, the second component was taken to be a black hole.
As remarked in Section 6, we have found that the coalescence for
tidally locked binaries (papers I, II) was less violent than
that for initially non-spinning polytropes (papers III, IV)---for
a stiff polytrope and a binary with $q<1$, the tidally locked polytrope
dribbled mass at discrete intervals, while the irrotational
polytrope was almost completely tidally disrupted in a single episode
of mass transfer.

Since the mass relationship for a polytrope is
$d\log R/d\log M=(\Gamma-2)/(3\Gamma-4)$,
a stiff polytrope ($\Gamma>2$) responds to mass loss by shrinking,
while the soft ones ($\Gamma<2$) expand when losing mass.
In this sense, the soft polytropes (which were always completely
disrupted in their first approach to the black hole) are a better
model for neutron stars (for which $dR/dM<0$, e.g., Arnett and
Bowers 1977), while the stiff ones may be taken as an approximation
to quark stars. Quark matter is nearly incompressible, with a density
contrast less than a factor of five inside a quark star (Witten, 1984),
and for lower mass quark stars $M\propto R^3$ is a good
approximation (Alcock et al., 1986a). In general,
the volume of quark stars increases with their mass, and
for $P\propto \rho - \rho_0$ the effective polytropic index
goes to infinity as $\rho\rightarrow\rho_0$ (Haensel et al. 1986),
so the outer parts of the (bare) quark star behave like an extremely
stiff polytrope. We expected and found that the
 overall evolution in the coalescence of a quark star
is very similar to that observed for
black hole binaries with a stiff polytrope (paper III).
However, there are
important differences, notably in the amount of matter
ejected. 

\begin{figure}
\psfig{width=\textwidth,file=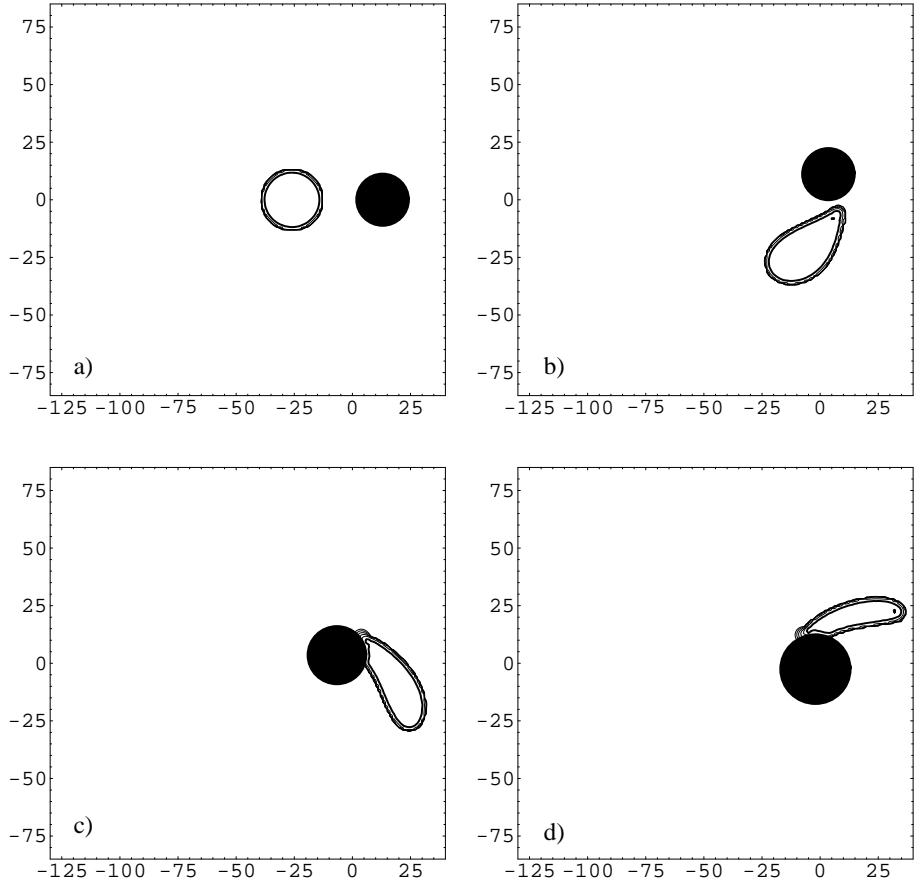,angle=0,clip=}
\caption{Density contours in the orbital plane at
(a)~$t=0$, (b)~$t=0.32\,$ms, (c)~$t=0.64\,$ms, (d)~$t=0.97\,$ms, 
(e)~$t=1.29\,$ms, (f)~$t=1.61\,$ms, (g)~$t=1.94\,$ms, and (h)~$t=2.26\,$ms,
for run C ($M_{\rm SS}=2.0 M_{\odot}$, $R_{\rm SS}=12.0\,$km). Orbital
rotation is counterclockwise, axes are labeled in km.
Contours are equally spaced every 0.5
dex, starting at $1.3\times10^{12}\,{\rm g\, cm}^{-3}$, with the
highest contour in bold at $\rho=\rho_0\equiv4.116\times10^{14}$~g~cm$^{-3}$.
Initially, as in panel a),
 the contours outside this bold contour are a numerical
artifact reflecting the size of the SPH kernel on the surface of
the quark star. At later stages, the thin contours represent average density
of quark ``dust'' composed of particles of unresolved mass,
less than $64M_{\rm SS}/17256$.
}
\end{figure}

\begin{figure}
\setcounter{figure}{0}
\psfig{width=\textwidth,file=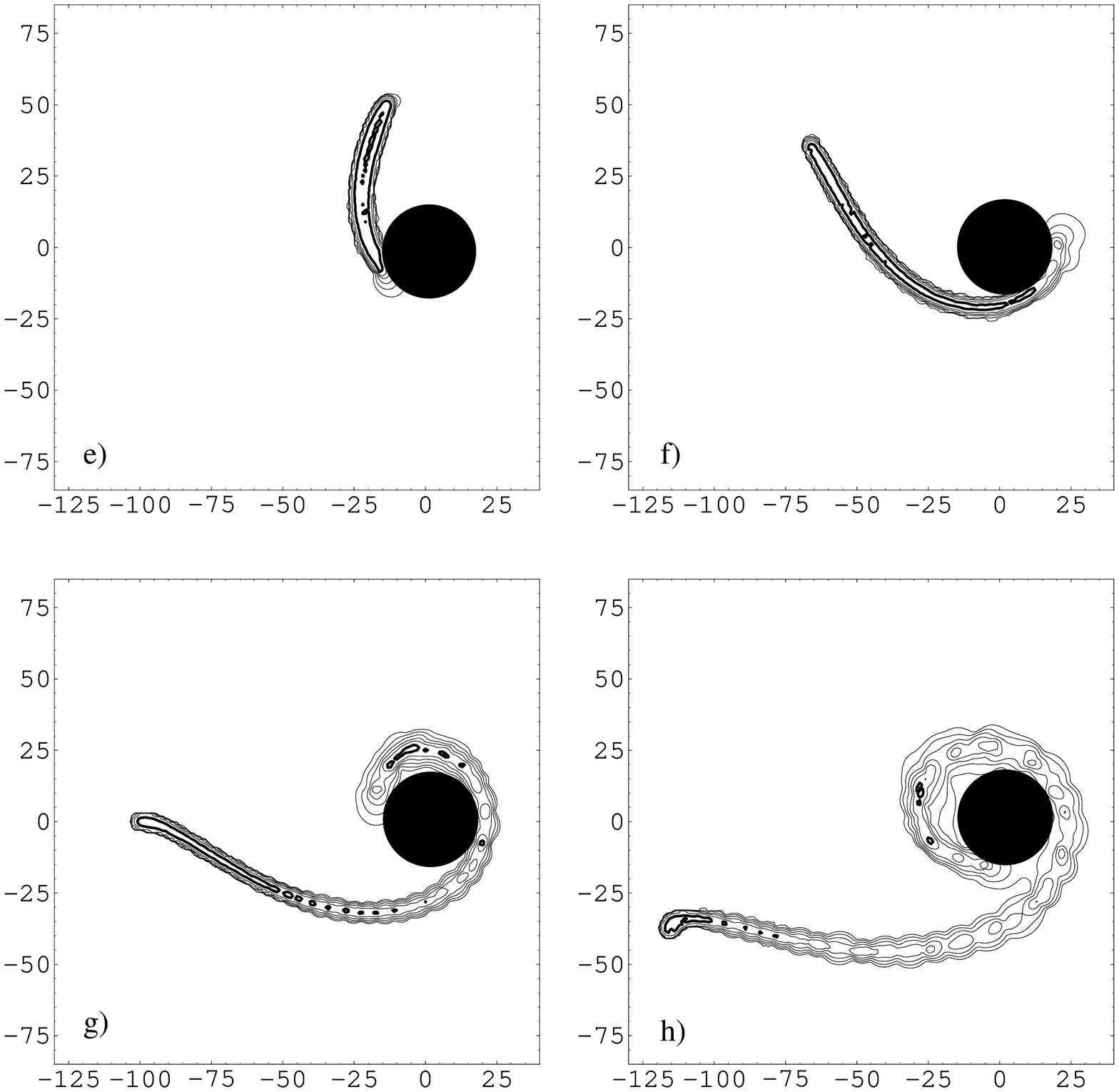,angle=0,clip=}
\caption{continued. Note the appearance of a large ``vacuum''
region of density below $\rho_0$ in the middle of the very elongated
quark star in panel e) (bold contour within a bold contour),
 and smaller such regions in panels b), d), and
f). In panels g) and h), in addition to dust, several blobs of
quark fluid on their way to the black hole are clearly visible.
One such blob is seen to be  separating from the tubular quark star
in panel f).
}
\end{figure}

\begin{figure}
\psfig{width=\textwidth,file=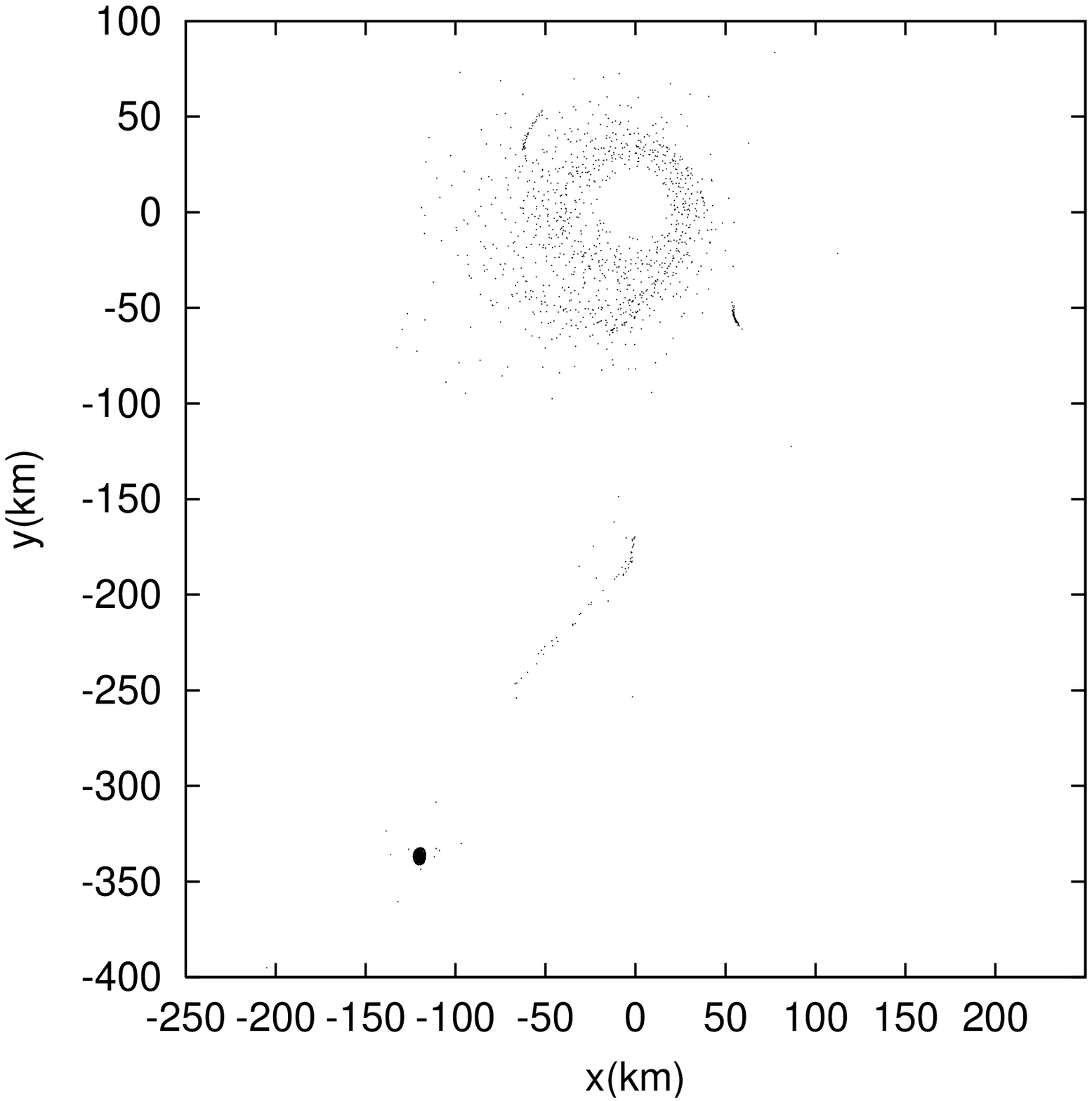,angle=0,clip=}
\caption{SPH particle positions at the end of the calculation for run
C (at $t=6.46$ ms), projected onto the orbital plane.  With the exception
of the clearly visible  clump at the end of the
disrupted tidal tail, the as-yet-not-accreted
quark matter has dispersed into droplets of
 unresolved mass. All SPH particles (dots in this figure)
have the same mass, $m_{i}=M_{\rm SS}/N$, so the mass density 
in the zero--pressure ``fog'' of 
these droplets
is proportional to the number of SPH particles per
unit volume. Everything visible in the figure, save at most two particles,
will eventually be accreted by the black hole.
}
\end{figure}

\begin{figure}
\psfig{width=\textwidth,file=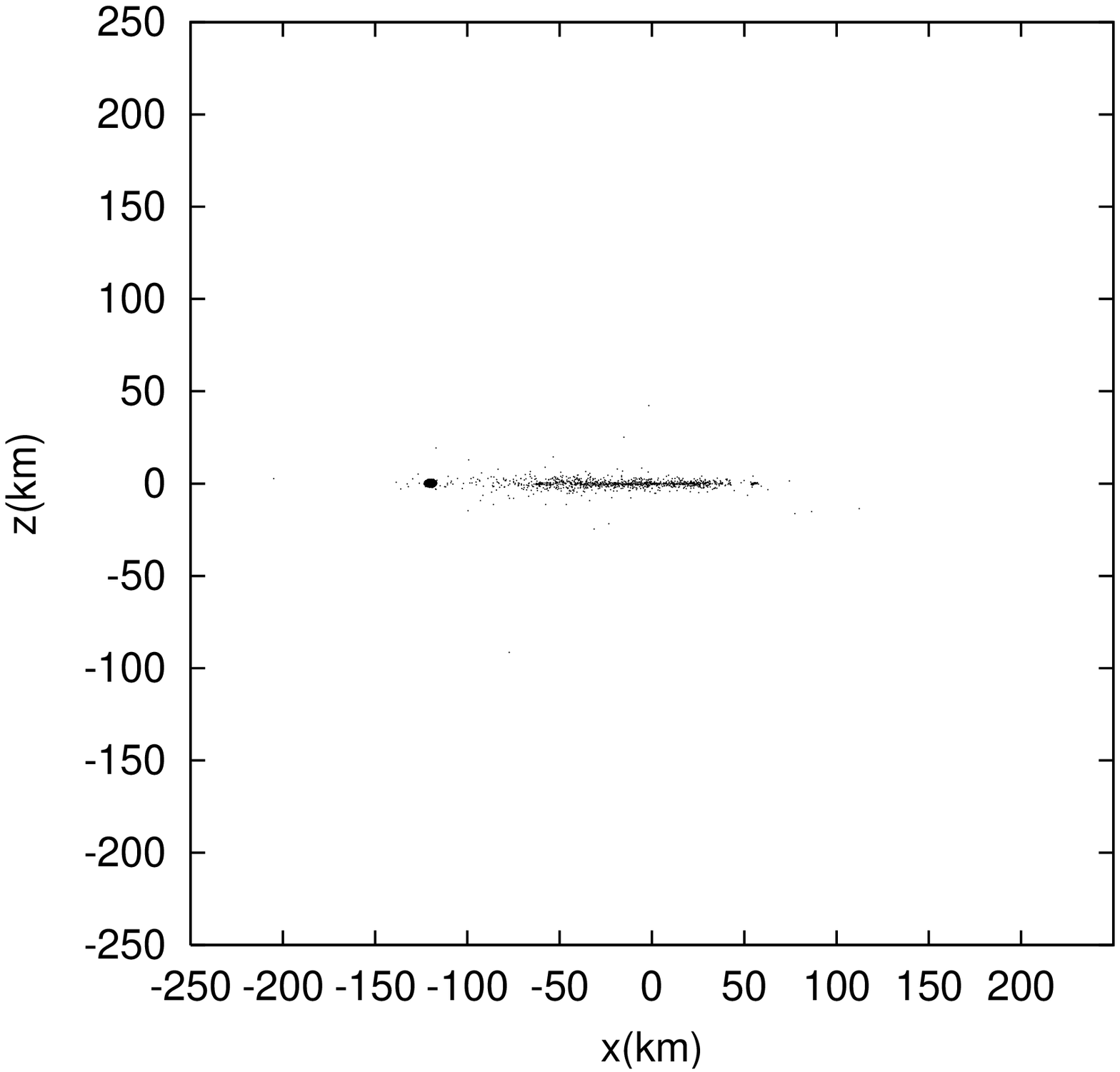,angle=0,clip=}
\caption{SPH particle positions at the end of the calculation for run
C (at $t=6.46$ ms), projected onto the meridional plane $y=0$. 
}
\end{figure}

\begin{figure}
\psfig{width=\textwidth,file=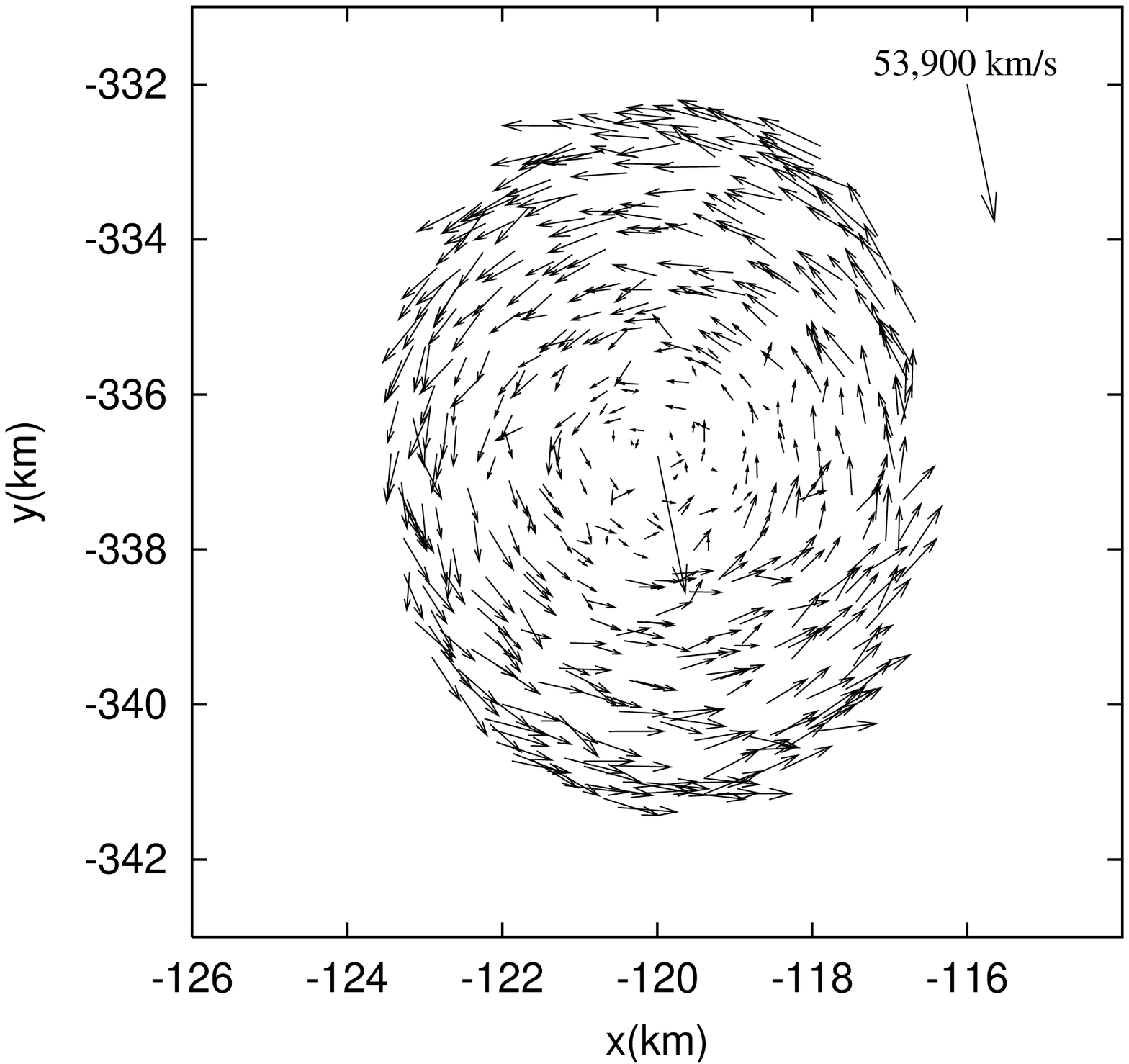,angle=0,clip=}
\caption{Relative velocity field for the clump at the end of 
run C, projected onto the orbital plane (compare Fig. 2). The arrow
in the top--right corner shows the clump center--of--mass velocity vector
(which has been subtracted from individual particle velocities
to show the rotational motion).
 The mass of this volume of quark fluid orbiting the black hole
is $0.055M_{\odot}$, and the inferred
angular velocity of rotation is
 $\Omega_{\rm rot}\approx 5\times10^3\,$s$^{-1}$.}
\end{figure}

\section{Results}

All our computations (regardless of the initial
mass $M_{\rm SS}$ and mass ratio~$q$) give qualitatively similar results.
 The
star becomes quickly elongated due to tidal forces,
 initiating mass transfer onto the black hole within one
orbital period. As the accretion stream winds around the black hole,
it forms a disk--like structure around it, while the portions of the
star farthest from the black hole form a tail of practically uniform
thickness.

The density in the fluid tidally stripped off the star
drops relatively quickly (within one orbital
period for the bulk of the matter) below the threshold value
$\rho_{0}$, so that in effect the fluid condenses into
a pressure--free fog, or dust
moving on ballistic trajectories which are practically determined by
the potential of the black hole.
 We see none of the expansion
of stripped matter characteristic of a polytrope.
This is simply because, in a sense, quark matter behaves more like
 a liquid than a gas---the volume of a mass $m$ of quark fluid cannot
exceed $m/\rho_0$, a fact also evident in the behavior of the star
itself, in which
already in the initial stages the pressure
decreases so quickly that the density {\it inside} the star drops
below $\rho_0$ and cavitation occurs, see Fig.~1, especially panel~e).
This last result may not be robust: the initial
central density in our Newtonian star is less than $3\rho_0$, so it
takes less than a threefold increase of its volume for cavitation to occur.
But when the equilibrium structure is governed by the relativistic
TOV equation, the central density is closer to $5\rho_0$ (Witten, 1984),
and a larger increase of volume can be accommodated.

Typically, a few high density ($\rho>\rho_0$) clumps break off the
star or condense out of the accretion stream. Most of these are
quickly accreted by the black hole. At the conclusion of tidal stripping
a starlet of $\sim 10^{-2}M_\odot$ remains relatively far from the black
hole, and on occasion is tidally injected into a highly elliptic
orbit---such is the case in run C, where the $0.055M_\odot$ 
starlet visible in Figs.~2
and~3 is in a bound orbit and still moving away from the black hole 
with orbital speed of $5.4\times10^4\,$km/s at the
end of the simulation (see Fig.~4 for the center-of-mass velocity vector).
The same tidal interaction has also substantially spun up the starlet
to a rotational period of 1.3 ms (at the end of simulation C). 

In the six runs presented here, the amount of mass ejected remains
unresolved. At most, a few
individual SPH particles (of mass $M_{\rm SS}/N$
each, see Table~1) are in unbound trajectories---the number of SPH
particles ejected  varies from none (zero) to three for the runs
of Table~1. For instance, at the end of run C only one SPH particle
(to the left of the $x=-200\,$km tick-mark in Fig.~2,
 at $y\approx -390\,$km) is on a clearly outbound trajectory,
its terminal velocity (at infinity) will be about $45 000\,$km/s,
and its velocity is so high that it
will not only leave the erstwhile binary, but the Galaxy and the Virgo
cluster as well.
The future of one more particle is undecided, it may or may not be
bound to the starlet, whose fate is sealed. The starlet is doomed to
undergo a close encounter with the black hole.

\begin{figure}
\psfig{width=\textwidth,file=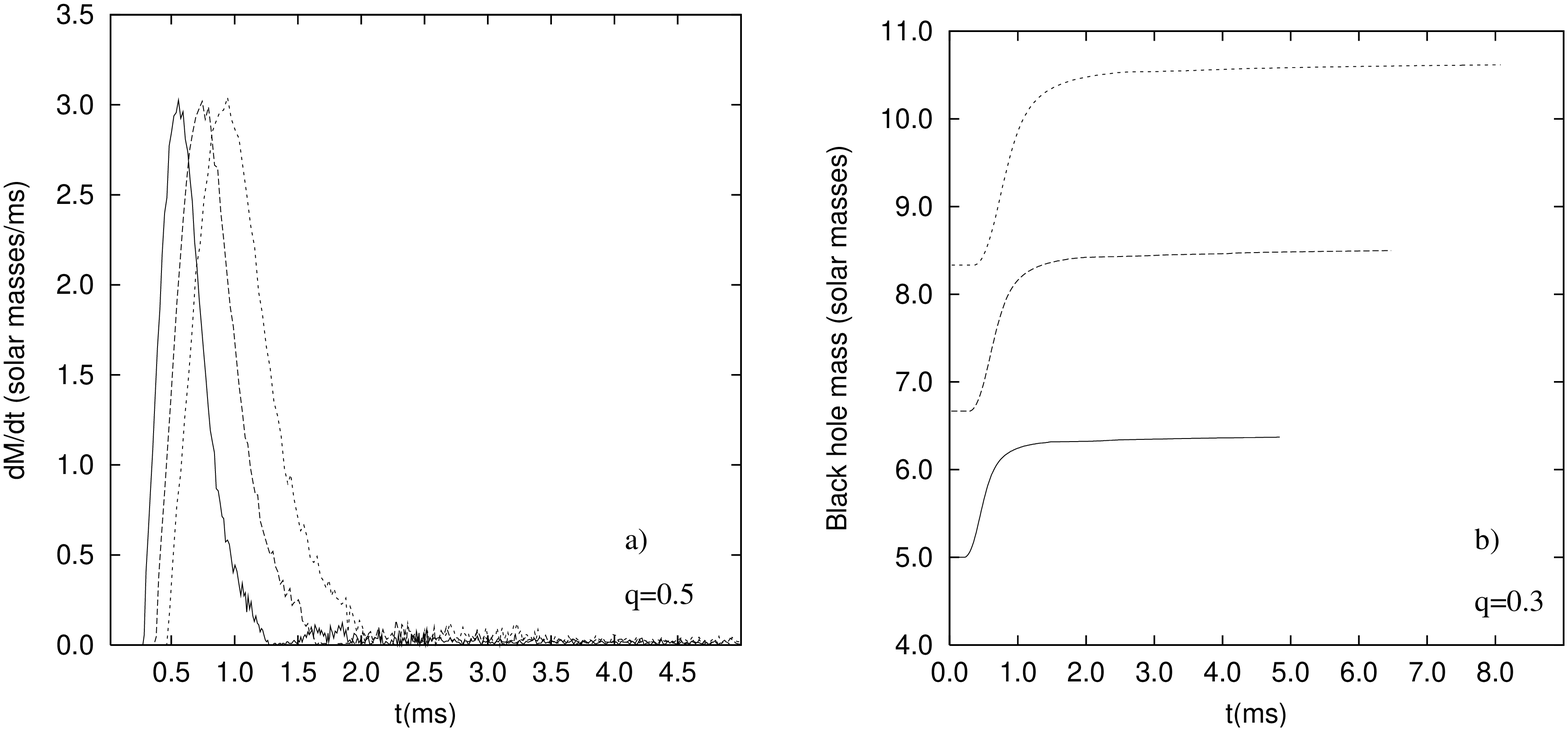,angle=0,clip=}
\caption{(a) Mass accretion rate onto the black hole as a function of
time for runs A (solid line), C (dashed line) and E (dotted line).
Accretion of individual ``condensations'' is clearly visible for
run A starting at about 1.5 ms. 
(b) Black hole mass
as a function of time for runs B (solid line), D (dashed line) and 
F(dotted line).
}
\end{figure}

\section{Discussion}
It has long been noted that the tidal action of a black hole on
a nearly incompressible star will stretch the latter into a long tube
(Wheeler 1971). Mashhoon (1973) computes the shape of a uniform-density
star approaching a Kerr black hole from infinity, and finds
(neglecting accretion and forcing the star to be a tri-axial ellipsoid)
that the star continues to be stretched even after passing the
point of closest approach to the black hole (periholon?), at
$1.961 GM_{\rm BH}/c^2$ for a particular choice of initial conditions,
reaching a seven-fold elongation at a distance of $5.5GM_{\rm BH}/c^2$.
It is gratifying to note that our Newtonian simulation gives qualitatively
similar results.

As in the case of stiff polytropes, we find that a small fragment of
the star survives the first encounter and is placed into a higher
orbit (papers~I, III, Lee and Klu\'zniak 1995, Klu\'zniak and Lee
1998). In all cases (Table~1), we find that the mass of this quark starlet
is $M_{\rm s}=0.03M_{\rm SS}$. However, gravitational radiation
will quickly lead to the coalescence with the black hole also of
this starlet. The equilibrium structure of such low-mass quark stars
is well described (at least up to the Jacobi turn-off) by
Maclaurin spheroids (Amsterdamski et al. 2002), but at the end of the
simulation the starlet is not yet in equilibrium. We will not speculate
here on the astrophysical signatures of such an object, as we do not
expect a similar result to hold for the Paczy\'nski-Wiita potential.
However, we do note that the starlet has been spun up by tidal interactions,
this had not been previously recognized.

We see no evidence of ``figure 8'' trajectories envisaged by 
Caldwell and Friedman (1991), and hence none of frequent and high-speed
collisions between stellar fragments, but we do confirm their estimates
of tidal fragmentation of the star to the limits of our resolution.
With the exception of the starlet,
the surviving fragments are definitely $<10^{-2}M_\odot$ in mass,
as the code would recognize any fragment of density exceeding
$\rho_0$ and mass not less than $64 M_{\rm SS}/N$
(each SPH particle
has 64 ``neighbors'' in the current implementation); in fact,
some even smaller 
high density fragments
have been resolved, e.g., the soon--to--be--accreted blobs
 visible in panels g) and h) of Fig. 1.
Paczy\'nski (1991) suggests the formation of an accretion disk,
and we do see a disk-like structure, but it is composed of small
fragments of (quark) matter, rather like Saturn's rings or Kuiper's belt.
At least, such is the case at the end of our simulation.
Again, this is related to the existence of a minimum
density, $\rho_0$, of quark matter (Section 8), there simply
is not enough fluid to fill the volume of a disk about $50\,$km
in radius and a few kilometers thick (Figs. 2 and 3).
Qualitatively, then, our simulation is in agreement with the
relativistic calculations of Wheeler (1971) and Mashhoon (1973),
and we find that some, but not all, of the expectations of
Caldwell and Friedman (1991) and of Paczy\'nski (1991) are supported
by our study.

An unexpected result is the very rapid accretion of much of the material
(Fig.~5). Most of the quark star has been devoured by the black hole
in less than 2 ms. Inclusion of a pseudo-Newtonian potential will make this
process even more drastic.

Finally, we find no definitive evidence of mass ejection from the system.
With our resolution we place an upper bound of $<3\times10^{-4}M_\odot$
on the amount of matter ejected, in all the cases considered.
This is in stark contrast to the polytropic case (papers I, III),
and is clearly a consequence of the equation of state of quark matter.
Our limit on the mass ejected is lower than that allowed 
by the work of Lattimer and Schramm (1974) for neutron
stars coalescing with black holes, but we have to agree with them
that ``the possibility of zero-mass ejection cannot be totally excluded.''

\section{Conclusions}

We have found no convincing evidence of ejection of quark matter
from the binaries modeled. Further simulations,
not reported here, show that inclusion of pseudo-Newtonian
potentials only strengthens this conclusion. This may encourage other
workers to continue studying the astrophysics of quark stars.

\section{Acknowledgements}

We gratefully acknowledge financial support from DGAPA--UNAM
(IN-110600) and CONACYT (27987E) and KBN (grant 2P03D00418).
WK wishes to acknowledge the hospitality of UNAM. 
It is a pleasure to thank Dany Page for many conversations.
We also thank the anonymous referee for helpful comments.

\end{document}